\documentclass[usenatbib,a4paper]{mn2e}

\usepackage{epsfig,times,amsmath,supertabular,colortbl,amsfonts,amssymb,multirow,array,fleqn,longtable}

\title[Galaxy-Galaxy Flexion Statistics]{Measuring Dark Matter Substructure with Galaxy-Galaxy Flexion Statistics}

\author[Bacon et al]{D. J. Bacon$^1$, A. Amara$^2$ \& J. I. Read$^{3,4}$\\ 
$^1$ Institute of Cosmology and Gravitation, University of Portsmouth,
Dennis Sciama Building, Burnaby Road, Portsmouth, PO1 3FX, United Kingdom\\
$^2$ Department of Physics, ETH Zurich, Wolfgang-Pauli-Strasse 16, CH-8093 Zurich, Switzerland\\
$^3$ Institute for Theoretical Physics, University of Zurich, Winterthurerstrasse 190 8047\\
$^4$ Department of Physics and Astronomy, University of Leicester, University Road, Leicester, LE1 7RH}

\begin{document} 
\maketitle

\newcommand{\expect}[1]{\left\langle #1 \right\rangle} 

\newcommand{\fflex}{\mbox{$\mathcal{F}$}}
\newcommand{\fflexsmall}{\mbox{\tiny $\mathcal{F}$}}
\newcommand{\gflex}{\mbox{$\mathcal{G}$}}
\newcommand{\gflexsmall}{\mbox{\tiny $\mathcal{G}$}}
\newcommand{\fflext}{\mbox{$\mathcal{F_T}$}}
\newcommand{\gflext}{\mbox{$\mathcal{G_T}$}}
\newcommand{\stild}{\tilde{\gamma}}
\newcommand{\ftild}{\tilde{\mbox{$\mathcal{F}$}}}
\newcommand{\gtild}{\tilde{\mbox{$\mathcal{G}$}}}
\newcommand{\mR}{\mbox{$\mathcal{R}$}}
\newcommand{\trans}{\mbox{$\mathcal{T}$}}
\newcommand{\tri}{\mathcal{T}}
\newcommand{\dif}{\mbox{$\mathrm{d}$}}
\newcommand{\x}{\mbox{\boldmath$x$}}
\newcommand{\y}{\mbox{\boldmath$y$}}
\newcommand{\thetab}{\mbox{\boldmath$\theta$}}
\newcommand{\Delthetab}{\mbox{\boldmath$\Delta\theta$}}
\newcommand{\alphab}{\mbox{\boldmath$\alpha$}}
\newcommand{\betab}{\mbox{\boldmath$\beta$}}
\newcommand{\xib}{\mbox{\boldmath$\xi$}}
\newcommand{\nablab}{\mbox{\boldmath$\nabla$}}
\newcommand{\me}{\mbox{$\mathrm{e}$}}
\newcommand{\mi}{\mbox{$\mathrm{i}$}}
\newcommand{\mes}{\mbox{\scriptsize$\mathrm{e}$}}
\newcommand{\mos}{\mbox{\scriptsize$\mathrm{o}$}}
\newcommand{\mess}{\mbox{\tiny$\mathrm{e}$}}
\newcommand{\moss}{\mbox{\tiny$\mathrm{o}$}}
\newcommand{\msep}{\mbox{\scriptsize$\mathrm{sep}$}}
\newcommand{\mis}{\mbox{\scriptsize$\mathrm{i}$}}
\newcommand{\ml}{\mbox{\scriptsize$\mathrm{l}$}}
\newcommand{\mls}{\mbox{\scriptsize$\mathrm{ls}$}}
\newcommand{\ms}{\mbox{\scriptsize$\mathrm{s}$}}
\newcommand{\mE}{\mbox{\scriptsize$\mathrm{E}$}}
\newcommand{\mB}{\mbox{\scriptsize$\mathrm{B}$}}
\newcommand{\mc}{\mbox{\scriptsize$\mathrm{c}$}}
\newcommand{\mH}{\mbox{$\mathrm{H}$}}
\newcommand{\mkms}{\mbox{$\mathrm{kms}$}}
\newcommand{\mkg}{\mbox{$\mathrm{kg}$}}
\newcommand{\mMpc}{\mbox{$\mathrm{Mpc}$}}
\newcommand{\mcov}{\mbox{$\mathrm{cov}$}}
\newcommand{\mN}{\mbox{\scriptsize$\mathrm{N}$}}
\newcommand{\mint}{\mbox{\scriptsize$\mathrm{int}$}}
\newcommand{\mcrit}{\mbox{\scriptsize$\mathrm{crit}$}}
\newcommand{\mtot}{\mbox{\scriptsize$\mathrm{tot}$}}
\newcommand{\mrot}{\mbox{\scriptsize$\mathrm{rot}$}}
\newcommand{\mprop}{\mbox{\scriptsize$\mathrm{prop}$}}
\newcommand{\mcom}{\mbox{\scriptsize$\mathrm{com}$}}
\newcommand{\mang}{\mbox{\scriptsize$\mathrm{ang}$}}
\newcommand{\mlum}{\mbox{\scriptsize$\mathrm{lum}$}}
\newcommand{\mvir}{\mbox{\scriptsize$\mathrm{vir}$}}
\newcommand{\mstar}{\mbox{\scriptsize$\mathrm{star}$}}
\newcommand{\mobs}{\mbox{\scriptsize$\mathrm{obs}$}}
\newcommand{\mMC}{\mbox{\scriptsize$\mathrm{MC}$}}
\newcommand{\mMCs}{\mbox{\tiny$\mathrm{MC}$}}
\newcommand{\atanh}{\mbox{$\mathrm{tanh}$}}
\newcommand{\nn}{\nonumber \\}
\newcommand{\sextractor}{\textsc{SExtractor}}
\newcommand{\mmin}{\mbox{\scriptsize$\mathrm{min}$}}
\newcommand{\mmax}{\mbox{\scriptsize$\mathrm{max}$}}
\newcommand{\munweighted}{\mbox{\scriptsize$\mathrm{unweighted}$}}
\newcommand{\mGaussian}{\mbox{\scriptsize$\mathrm{Gauss}$}}
\newcommand{\mdiag}{\mbox{\scriptsize$\mathrm{diag}$}}
\newcommand{\mmag}{\mbox{\scriptsize$\mathrm{unweighted}$}}
\newcommand{\mav}{\mbox{\scriptsize$\mathrm{av}$}}
\newcommand{\mshot}{\mbox{\scriptsize$\mathrm{shot}$}}
\newcommand{\mlin}{\mbox{\scriptsize$\mathrm{lin}$}}
\newcommand{\mback}{\mbox{\scriptsize$\mathrm{back}$}}
\newcommand{\mcoll}{\mbox{\scriptsize$\mathrm{coll}$}}
\newcommand{\mf}{\mbox{\scriptsize$\mathrm{f}$}}
\newcommand{\mhalos}{\mbox{\scriptsize$\mathrm{halos}$}}
\newcommand{\mm}{\mbox{\scriptsize$\mathrm{m}$}}
\newcommand{\lcdm}{$\Lambda$CDM}
\newcommand{\ard}{\hat{a}^{\dagger}_r}
\newcommand{\ald}{\hat{a}^{\dagger}_l}
\newcommand{\ards}{\hat{a}^{\dagger 2}_r}
\newcommand{\alds}{\hat{a}^{\dagger 2}_l}
\newcommand{\ar}{\hat{a}_r}
\newcommand{\al}{\hat{a}_l}
\newcommand{\ars}{\hat{a}^2_r}
\newcommand{\als}{\hat{a}^2_l}

\newcommand{\nat}{Nat}
\newcommand{\mnras}{MNRAS}
\newcommand{\apj}{ApJ}
\newcommand{\apjl}{ApJL}
\newcommand{\apjs}{ApJS}
\newcommand{\physrep}{Phys.~Rep.}
\newcommand{\aap}{A\&A}
\newcommand{\aaps}{A\&AS}
\newcommand{\aj}{AJ}
\newcommand{\prd}{Phys.~Rev.~D.}
\def\com#1{{\bf \large [#1]}}

\begin{abstract}
It is of great interest to measure the properties of substructures in dark matter halos at galactic and cluster scales. Here we suggest a method to constrain substructure properties using the variance of weak gravitational flexion in a galaxy-galaxy lensing context. We show the effectiveness of flexion variance in measuring substructures in N-body simulations of dark matter halos, and present the expected galaxy-galaxy lensing signals. We show the insensitivity of the method to the overall galaxy halo mass, and predict the method's signal-to-noise for a space-based all-sky survey, showing that the presence of substructure down to $10^9 M_\odot$ halos can be reliably detected. 
\end{abstract}


\section{Introduction}

The amount of substructure in dark matter halos on galactic and cluster scales is a question of considerable interest. $\Lambda$CDM cosmological simulations predict thousands of dark satellites within the virial radius of the Milky Way, yet to date only $\sim 20$ have been observed \citep[e.g.][]{1993MNRAS.264..201K, 1999ApJ...522...82K, 2007ApJ...667..859D}. Similar results have been obtained for our nearest neighbour Andromeda \citep{2006MNRAS.365..902M}, and in galaxy groups \citep{2007arXiv0704.2604D}. Averaging over galaxies observed in the field, there appears to be a suppression in the expected number below a baryonic mass of $\sim 10^{10}$\,M$_\odot$ \citep[e.g.][]{2005RSPTA.363.2693R}. This dearth of low mass galaxies could be telling us that galaxy formation becomes increasingly inefficient below the peak of the luminosity function \citep[e.g.][]{1986ApJ...303...39D, 2000ApJ...539..517B, 2002MNRAS.333..177B, 2004ApJ...609..482K, 2006MNRAS.371..885R}. Alternatively, it could be telling us something about the nature of dark matter \citep[e.g.][]{2001ApJ...556...93B, 2001ApJ...558..482B, 2001ApJ...559..516A, 2008MNRAS.386.1029K}, or about the details of inflation \citep{2003ApJ...598...49Z}.

Gravitational lensing is a powerful tool for probing substructure within galaxies and clusters. Unlike photometric galaxy surveys, lensing directly probes the dark Universe; in principle, even if completely dark subhalos exist, these will have some influence on the observed gravitational lensing signal and may therefore be detected. Several lensing techniques for constraining substructure already exist: one can examine the strong lensing of quasars by galaxies and clusters \citep[e.g.][]{1998MNRAS.295..587M,2001ApJ...563....9M,2002ApJ...567L...5M,2003ApJ...598..138K,2003MNRAS.339..607M,2004ApJ...610...69K,2004ApJ...607...43M, 2005ApJ...622...72M,2006MNRAS.372.1692A,2006MNRAS.367.1367A, 2007MNRAS.382.1225M,2008MNRAS.385.2107S} including the time delay phenomenon \citep{2009ApJ...699.1720K}. One can also constrain the substructure using weak galaxy-galaxy shear \citep{2004astro.ph.11426N}. Here we examine the usefulness for measuring substructure of another weak lensing phenomenon: flexion, building on several earlier studies \citep[c.f.][]{2002ApJ...564...65G,2005ApJ...619..741G, 2006MNRAS.365..414B, 2006ApJ...645...17I,2008A&A...485..363S}.

It has been noted by these authors that flexion responds to small-scale variations in the gravitational potential. In a galaxy-galaxy lensing context, the mean flexion in annuli around a halo will fall off rapidly, as the mean halo density gradient is small away from the central region. However, if substructure is present, the flexion {\em variance} in annuli may not fall off so quickly, as the substructures will lead to potential fluctuations which will cause a flexion varying rapidly from place to place. It is this idea that the current paper will examine.

The topic is of particular interest in the context of forthcoming and planned large lensing surveys. Ground-based surveys such as Pan-STARRS\footnote{http://pan-starrs.ifa.hawaii.edu} and the Dark Energy Survey\footnote{https://www.darkenergysurvey.org} will obtain many thousands of square degrees of lensing-quality data, allowing very precise galaxy-galaxy lensing constraints. Beyond this, space-based survey telescopes such as Euclid\footnote{http://www.ias.u-psud.fr/imEuclid} will obtain extraordinary accuracy for galaxy-galaxy lensing due to a high number density of galaxies and a survey area of 20000 square degrees. This paper will make some initial predictions for the level of accuracy of flexion variance measurements with the latter survey.

The paper is organised as follows. In section \ref{Theory}, we consider the relevant theory of weak gravitational lensing, concentrating on the flexion formalism. We proceed to define the galaxy-galaxy flexion statistics that we require, and show how they can be estimated theoretically. In section \ref{Simulations} we describe N-body simulations of halos with and without substructure, and explain the procedure for creating lensing maps from these simulations. In section \ref{Analysis} we use the maps to illustrate the sensitivity of flexion to substructure, then calculate the expected galaxy-galaxy flexion variance signal together with its signal to noise for a Euclid-like survey. We examine the effect of varying the dominant halo mass, finding little impact, and show the much more pronounced effect of varying the amount of substructure present. This latter finding will illustrate the usefulness of the technique as a probe of dark matter temperature. We present our conclusions in section \ref{Conclusions}.

\section{Theory}
\label{Theory}

\subsection{Flexion}

The flexion formalism described here is more fully developed in \citet{2006MNRAS.365..414B}. We begin by noting that lensing in the weak regime can be described by mapping the surface brightness of a galaxy in the source plane, $f_S(\beta_i)$, to the surface brightness in the image plane, $f_I(\theta_i)$:
\begin{equation}
\label{eq:f_I}
f_I(\theta_i)=f_S(\beta_i)=f_S\left(A_{ij} \theta_j + \frac{1}{2}D_{ijk}\theta_j\theta_k\right).
\end{equation}
Here we have introduced several quantities: firstly, the Jacobian matrix $A$, which is taken to be constant across the galaxy image in the weak regime. It can be written in terms of lensing quantities
\begin{equation}
A=\left(\begin{array}{cc}1-\kappa&0\\0&1-\kappa\end{array}\right)+\left(\begin{array}{cc}-\gamma_1&-\gamma_2\\-\gamma_2&\gamma_1\end{array}\right)
\end{equation}
where $\kappa$ is the convergence; this maps a unit circle in the image plane to a circle with radius $1-\kappa$ in the source plane. In the case of an isolated lens, the convergence is proportional to the projected density of matter in the lens \citep[c.f.][]{2001PhR...340..291B},
\begin{equation}
\kappa(\vec{\theta})=\frac{\Sigma(D_l \vec{\theta})}{\Sigma_{cr}}
\end{equation}
where $\Sigma$ is the 2D projection of the density $\rho$,
\begin{equation}
\Sigma(\vec{\xi})=\int dr_3 \rho(\vec{\xi},r_3)
\end{equation}
where the integration is over the radial distance $r_3$ and $\Sigma_{cr}$ is given by
\begin{equation}
\Sigma_{cr}=\frac{c^2}{4\pi G}\frac{D_s}{D_l D_{ls}}
\end{equation}
where $D_l, D_s$ and $D_{ls}$ are the angular diameter distances from the observer to the lens, from the observer to the source, and from the lens to the source respectively.
If we also define the lensing potential $\psi$, which is proportional to the projection of the gravitational potential $\Phi$,
\begin{equation}
\psi(\vec{\theta})=\frac{4\pi G}{c^2}\frac{D_l D_s}{D_{ls}}\int dr_3 \Phi(D_l \vec{\theta}, r_3)
\end{equation}
then we can also write the convergence as
\begin{equation}
\kappa=\frac{1}{2}(\partial_1^2+\partial_2^2)\psi
\end{equation}
The other term in $A$ is the shear $\gamma_i$; this maps a circle in the image plane to an ellipse in the source plane. Its components can be written as
\begin{equation}
\gamma_1=\frac{1}{2}(\partial_1^2-\partial_2^2)\psi \,\,\,\,\,\,\,\,\,\, \gamma_2=\partial_1\partial_2\psi
\end{equation}
The next term in equation (\ref{eq:f_I}) is the $D$-tensor; this contains the lensing information at the next order of approximation, and corresponds to the varying of convergence and shear across an object. As shown in \citet{2006MNRAS.365..414B}, the $D$-tensor can be written in terms of the flexions,
\begin{eqnarray}
-2D_{ij1}&=&\left(\begin{array}{cc}3F_1&F_2\\F_2&F_1\end{array}\right)+\left(\begin{array}{cc}G_1&G_2\\G_2&-G_1\end{array}\right)\nonumber\\-2D_{ij2}&=&\left(\begin{array}{cc}F_2&F_1\\F_1&3F_2\end{array}\right)+\left(\begin{array}{cc}G_2&-G_1\\-G_1&-G_2\end{array}\right)
\label{eq:d}
\end{eqnarray}
where $G_i$ are the components of 3-flexion, describing the degree to which an object resembles a trefoil, and $F_i$ are the components of 1-flexion, describing the skewed shape of an object. We will only consider the 1-flexion for the purposes of this paper, which has the property of being the gradient of the convergence,
\begin{eqnarray}
F_1=\frac{1}{2}\left(\partial_1^3 + \partial_1\partial_2^2\right) \psi=\partial_1\kappa\nonumber\\
F_2=\frac{1}{2}\left(\partial_1^2\partial_2+\partial_2^3\right)\psi=\partial_2\kappa
\end{eqnarray}
It is this property of 1-flexion that is so important for the technique of this paper; 1-flexion will respond wherever the density is varying rapidly from place to place, which is the case with substructure.

\subsection{Galaxy-Galaxy Flexion}

We can now introduce the flexion variance in annuli as a substructure probe. It is usual in galaxy-galaxy lensing \citep[e.g.][]{2006A&A...455..441K, 2006MNRAS.368..715M} to choose particular galaxy samples which act as lens and source sets. The sets may have some members in common, depending on the technique used for lens-source correlation, but at any rate the selections can be achieved either via photometric redshifts, spectroscopic redshifts or a combination of the two.

Let us orientate ourselves with the most familiar case of galaxy-galaxy lensing, which involves shears. The angular separation $\theta$ of foreground object $f$ and background object $b$ is measured, and the shear of $b$ is decomposed into components tangential and diagonal to the line connecting $f$ and $b$,

\begin{equation}
\gamma_t=-\Re \left(\gamma {\rm e}^{-2i\phi}\right) \,\,\,\,\,\,\,\,\,\,\,\,
\gamma_B=-\Im \left(\gamma {\rm e}^{-2i\phi}\right)
\end{equation}
where $\phi$ is the position angle of the line, $\gamma=\gamma_1+i\gamma_2$, and $\Re$ and $\Im$ denote real and imaginary parts respectively. For a circular foreground lens, $\gamma_t$ will be activated by gravitational lensing, while $\gamma_B$ will not; if it is present, it is due to systematic effects.

Similarly a background flexion can be decomposed, but whereas shear is decomposed into tangential and diagonal components, flexion is decomposed into radial and tangential components due to the fact that it has different rotational properties to shear:
 
\begin{equation}
F_r=-\Re\left(F {\rm e}^{-i\phi}\right)\,\,\,\,\,\,\,\,\,\,\,\,
F_B=-\Im\left(F{\rm e}^{-i\phi}\right)
\end{equation}
where $F=F_1+iF_2$. Here the $r$ component is activated by gravity for a circular lens, while the $B$ component is unactivated unless systematics are present.

The behaviour of flexion at an angular distance $\theta$ from a singular isothermal sphere (SIS) was considered by \citet{2006MNRAS.365..414B}. They show that while the surface density of the SIS is proportional to $\theta^{-1}$, the  1-flexion and 3-flexion drop as $\theta^{-2}$. While real galaxies are not truly SIS, they are close enough to this profile that if we take foreground-background pairs, measure $F_r$, and consider the mean $F_r$ in annuli, we will see a similarly rapid drop in galaxy-galaxy flexion with annulus radius $\theta$.

This will remain true if there are substructures at angular distance $\theta$ from the centre of foreground galaxies. Although in the extreme locality there will be a larger flexion than usual, the mean signal averaged around the galaxy will still drop rapidly with $\theta$.

However, this will not be true for the flexion {\em variance} in the annulus. This will respond to any density fluctuations within the annulus. Therefore, in any annuli with non-negligible substructure, even if the mean flexion is small (as the mean gradient of density is small), the flexion variance will remain comparatively large. It is this behaviour that we will use to constrain substructure on galactic scales. 

At this point one may ask whether flexion variance is the best tool for our task; wouldn't flexion correlation functions in annuli provide more information? The question could be informed by experience in cosmic shear studies, where shear correlation functions provide more finesse than shear variance in cells. 

However, in our present case a correlation function does not seem to be helpful. Since the correlation function in question would be in annuli around a foreground galaxy, it constitutes a form of galaxy-galaxy-galaxy lensing \citep{2005A&A...432..783S}; it is a three-point statistic. In order to use this to measure substructures, it is necessary for two background sources to be close to the same substructure as well as to the foreground galaxy; this rarely happens, leading to low signal-to-noise. On the other hand, the flexion variance in annuli only involves two points, a foreground-background pair, with much greater signal-to-noise as we shall see.

\section{Simulations}
\label{Simulations}

In this section we describe N-body simulations which we will use to demonstrate the utility of flexion variance as a probe of substructure.

\subsection{3-D Density}

We use the cosmological $\Lambda$CDM simulation already presented in \citet{2005MNRAS.364..367D}. The simulation was run using PkdGRAV \citep{2001PhDT........21S}, with cosmological parameters: $(\Omega_{\rm m}, \Omega_{\Lambda}, \sigma_8, h) = (0.268, 0.732, 0.7, 0.71)$, and a box of size $L_{\rm box}=90$ Mpc, with $300^3$ particles. The initial conditions were generated with GRAFIC2 \citep{2001ApJS..137....1B}. From the simulation volume, we extracted four Milky Way sized halos at a mass resolution of $m_p=5.7\times 10^5 M_\odot$; their virial masses are $[2.1, 1.5, 1.2, 1.3]\times10^{12}M_\odot$. While we are therefore very limited in our number of lenses (to three projections of each of four high-resolution galaxies), we will find that this is sufficient to give the initial indicative results required by this paper.

\begin{figure}
\psfig{file=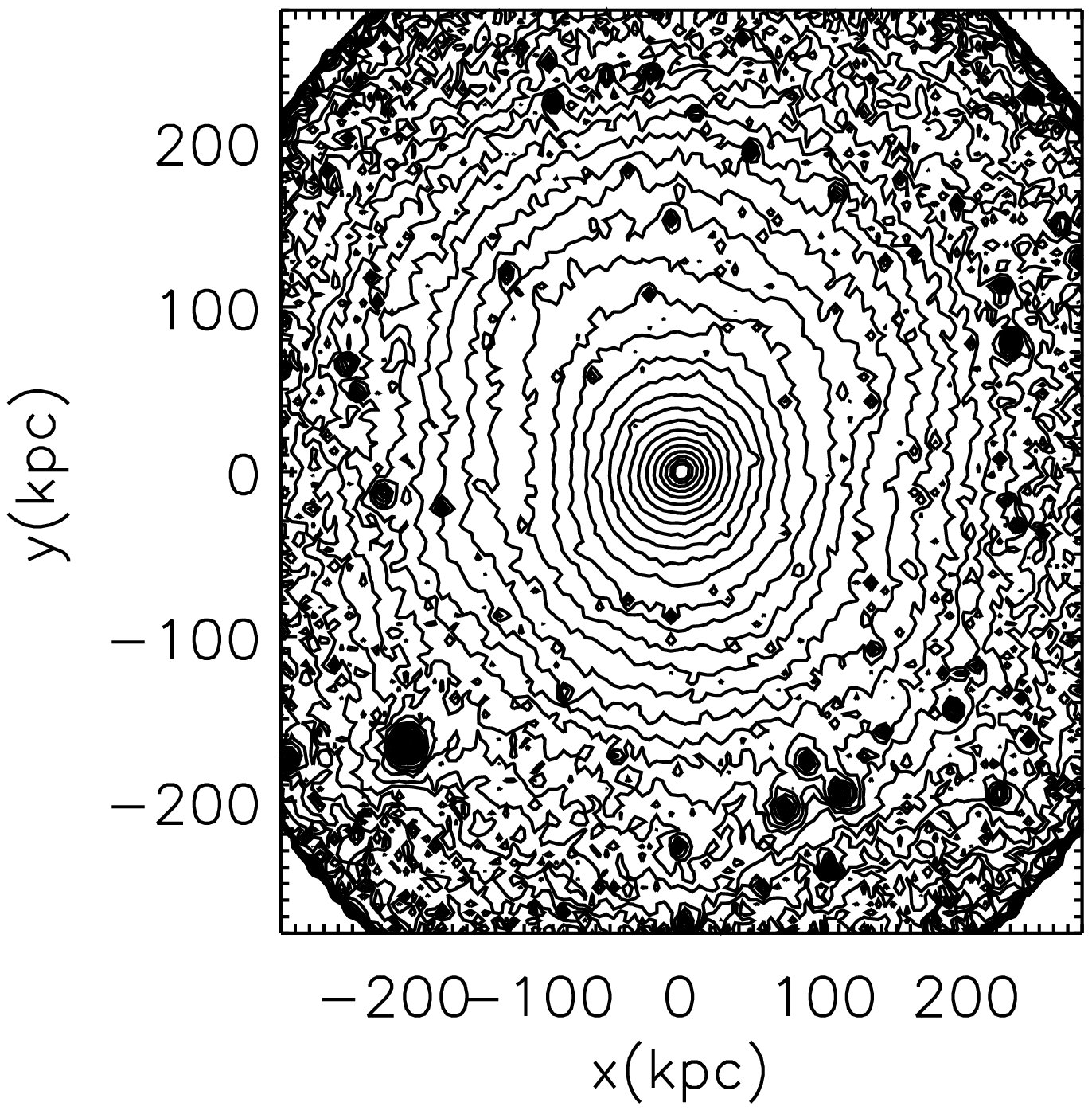,width=8cm}
\psfig{file=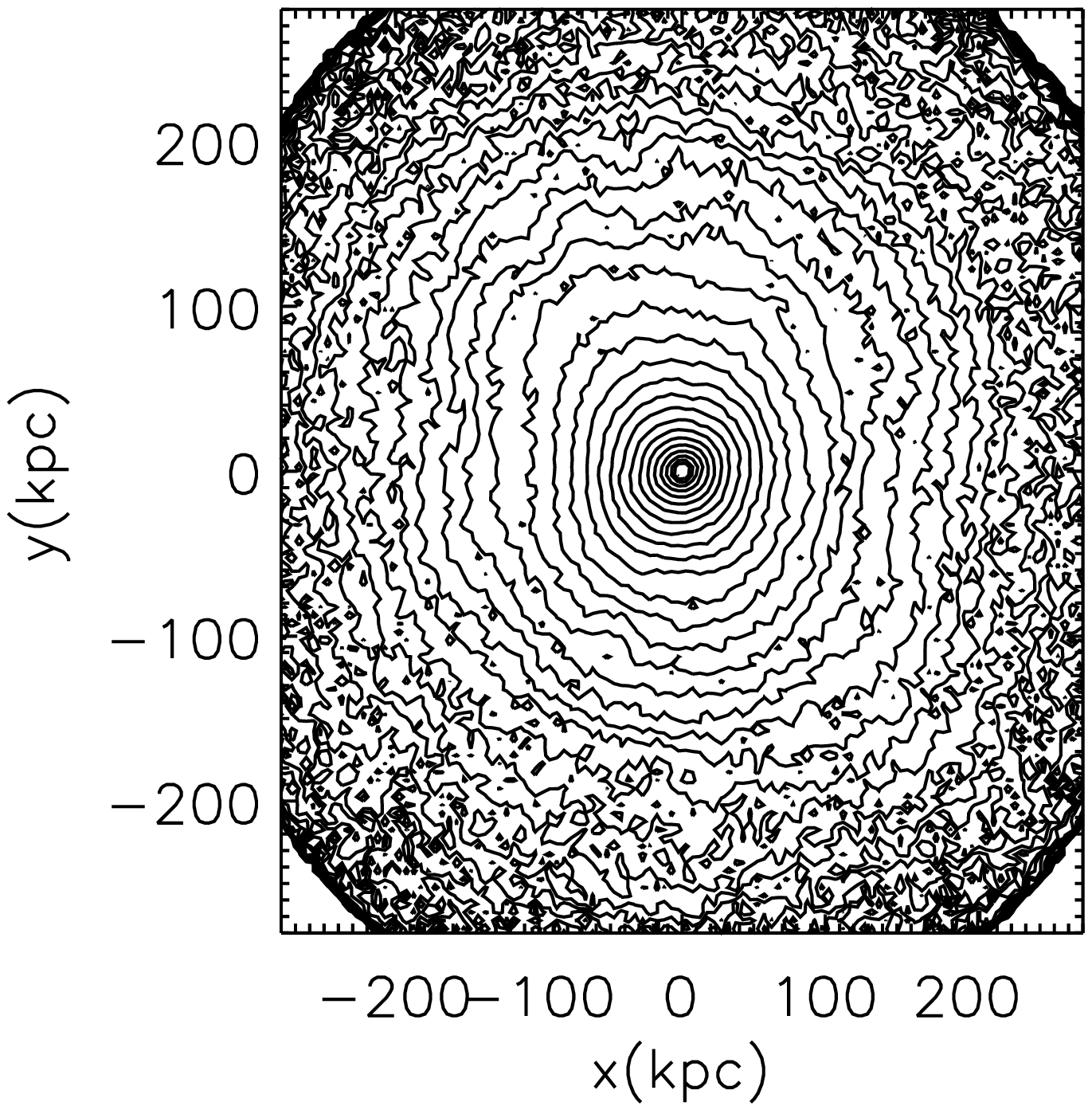,width=8cm}
\caption{N-body simulation of one of our Milky Way mass halos, with substructure (top panel) and without substructure (bottom panel), with contours showing projected logarithmic density over the range $[0.02,672] $\,M$_\odot$\,pc$^{-2}$.}
\label{fig:nbody}
\end{figure}

As in \citet{2008MNRAS.389.1041R}, the subhalos inside each `Milky Way' at redshift $z=0$ were identified using the {\it AHF} algorithm \citep{2004MNRAS.351..399G}. We considered all subhalos with $>50$ particles and assigned particles to the smallest structure they appear in so that each particle was counted only once. In some cases we will remove substructure; this is achieved by subtracting all particles not assigned to the main halo. An example halo, with and without substructure, is shown in Figure \ref{fig:nbody}.

\subsection{2-D Convergence}
The 3D numerical simulations discussed above represent the density field using discrete particles. We transform these into convergence maps by projecting the particles along particular spatial directions and placing the particles onto a 2D 1024x1024 grid, which we carry out using the IDL cloud-in-cell routine available as part of The IDL Astronomy User's Library\footnote{http://idlastro.gsfc.nasa.gov}. We produce three projection maps for each 3D halo (by projecting along the x, y or z axis). 

We investigated a number of techniques for filtering the mass maps. This is important because the finite number of simulation particles introduces shot noise into the 2D maps; this can compete with the substructure signal we are investigating. Here we show the results obtained using Multiscale Entropy Filtering (MEF) \citep{2006A&A...451.1139S}. This provides superior performance to a simple Guassian filter, as it reduces the shot noise while preserving the density fluctuations, as we shall demonstrate. For this purpose we use the routines supplied in the public code MRLens (Multi-Resolution methods for gravitational LENSing)\footnote{$\rm http://irfu.cea.fr/Phocea/Vie\_des\_labos/Ast/ast_visu.php?id\_ast=878$}, choosing the eight scale for MEF with first detection at the second scale. The 2D mass maps ($\Sigma$) are then rescaled using the critical density ($\Sigma_c$) to give the convergence ($\kappa$). We will use a lens-source configuration of $D_s=1200$ Mpc, $D_l=860$ Mpc, and $D_{ls}=610$ Mpc corresponding to our choice of median redshifts for foreground ($z=0.46$) and background $(z=1.1)$ sources appropriate for our projected Euclid survey. 

\section{Analysis}
\label{Analysis}

Armed with this set of simulations, we are in a position to examine whether substructure can be reliably measured using galaxy-galaxy flexion variance.

\subsection{Flexion Variance Sensitivity to Substructure}

\begin{figure}
\psfig{file=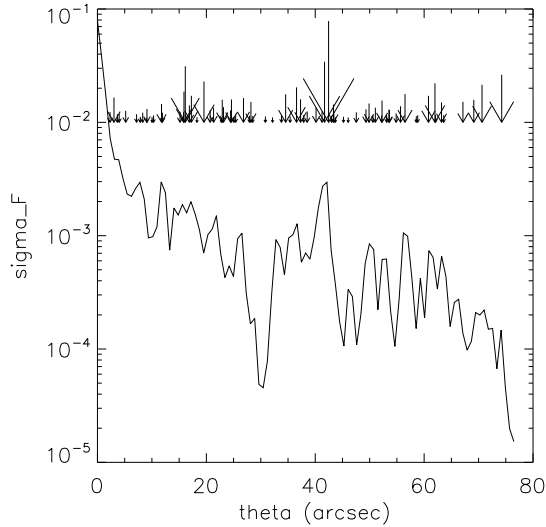,width=8cm}
\caption{Standard deviation of 1-flexion in annuli for one galaxy halo, in units of (arcsec)$^{-1}$. Arrows show the positions of subhalos, with arrow length proportional to subhalo mass.}
\label{fig:onehalo}
\end{figure}

Using the convergence maps for each halo, we calculate the related 1-flexion map using $F_i=\partial_i \kappa$, and smooth the resulting flexion map with a $0.6''$ diameter top hat filter to remove small-scale flexion peaks inaccessible to galaxy shape measurements. We then calculate the mean and standard deviation of $F$ in annuli with width $0.8''$, centred on the mode of the galaxy's $\kappa$ distribution. 

As an initial example, Figure \ref{fig:onehalo} shows the standard deviation $\sigma_F$ for the radial 1-flexion in annuli, for one example halo. Also displayed are arrows showing the radial positions of subhalos found in the simulation, with arrow size proportional to the mass of the subhalo. We can see that there is a rather close correspondence between the flexion variance and the subhalo positions and masses. This encourages us to examine what the signal will be for an ensemble of galaxy halos in a galaxy-galaxy lensing context.

\subsection{Galaxy-Galaxy Flexion Signal}

\label{signal}

\begin{figure}
\psfig{file=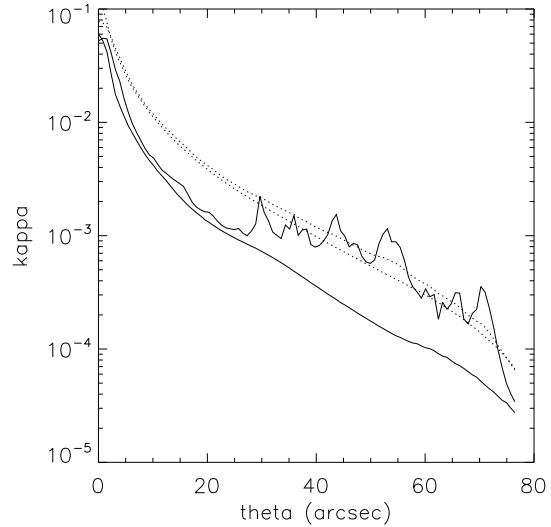,width=8cm}
\caption{Mean (dotted lines) and standard deviation (solid lines) of convergence $\kappa$ in annuli for the twelve halo orientations combined. Upper lines in each case show results with substructure, while lower lines show the results when substructure is removed.}
\label{fig:kapstat}
\end{figure}

\begin{figure}
\psfig{file=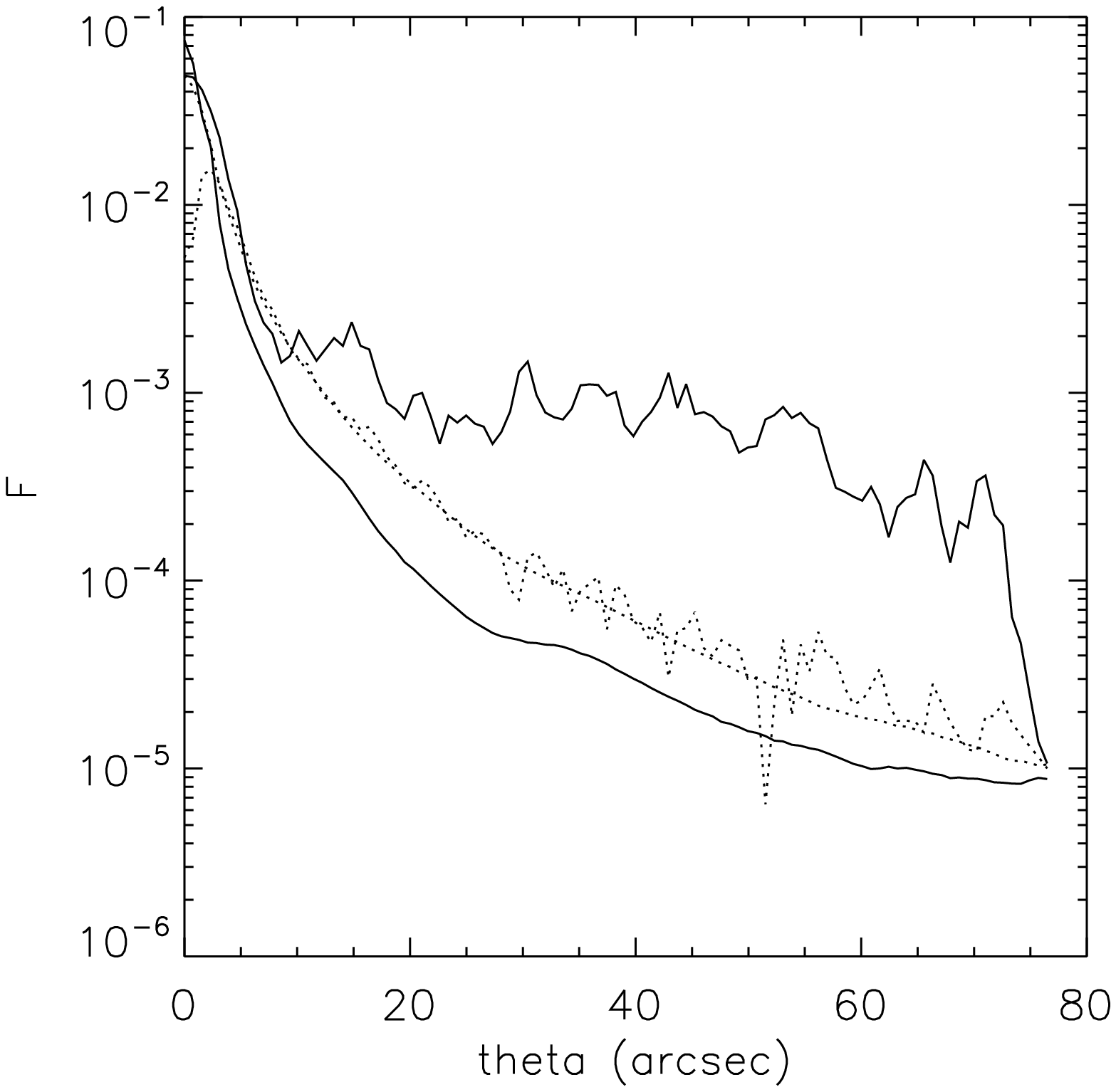,width=8cm}
\psfig{file=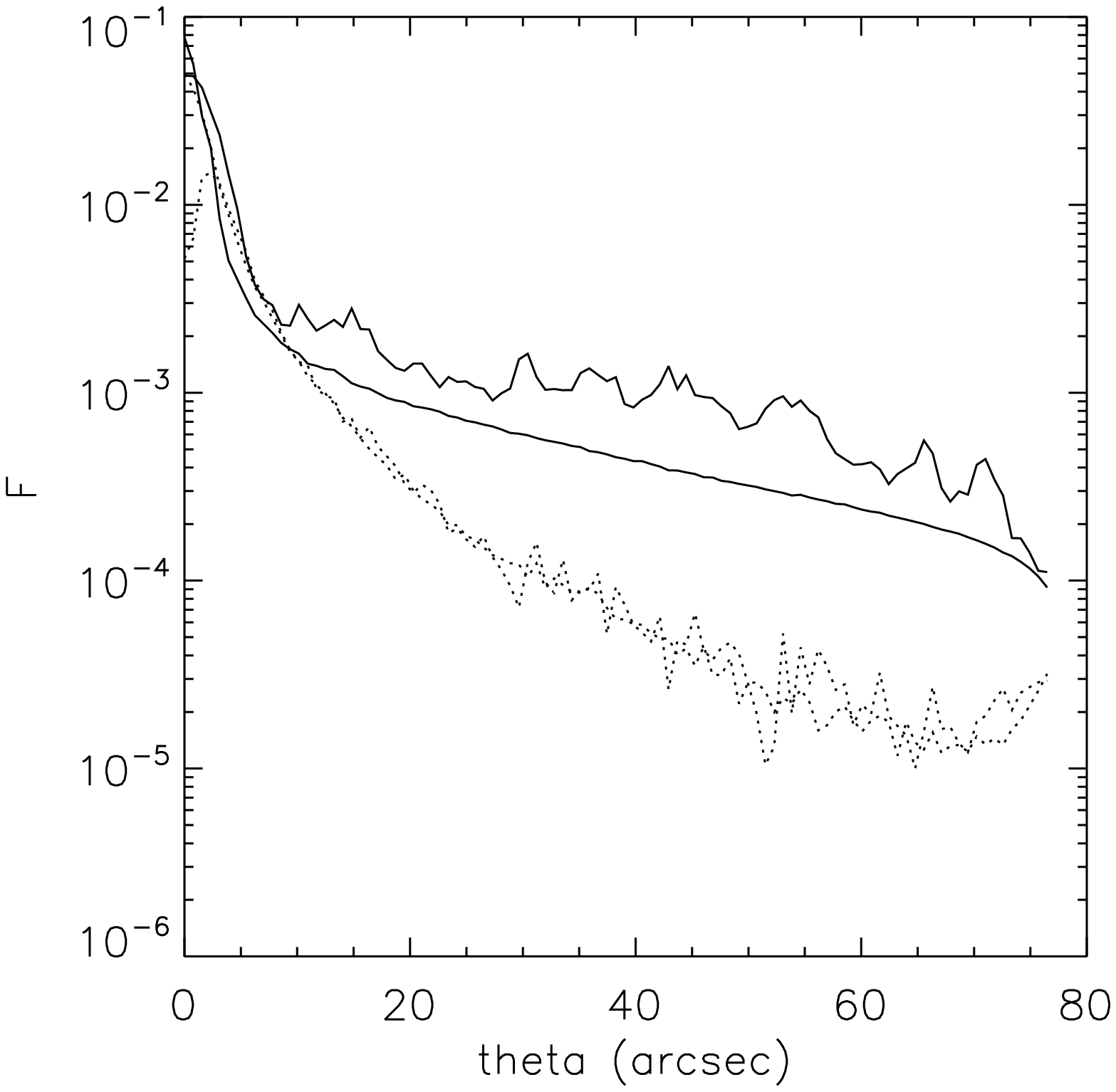,width=8cm}
\caption{Mean (dotted lines) and standard deviation (solid lines) of 1-flexion in annuli for the twelve halo orientations combined, in units of (arcsec)$^{-1}$. Upper lines in each case show results with substructure, while lower lines show results when substructure is removed. Top panel: MEF results; bottom panel: unfiltered results.}
\label{fig:fstats}
\end{figure}

In this section we will present an estimate of the galaxy-galaxy flexion signal; but in order to begin with a familiar quantity, we first examine the convergence for our halos in annuli of width $0.8''$, centred on the mode of the main halos' $\kappa$ distribution. Figure \ref{fig:kapstat} shows the mean convergence $\bar{\kappa}$ and standard deviation $\sigma_\kappa$  for an ensemble of background galaxies behind the twelve stacked halos, in annuli with radius $\theta$. Note that $\sigma_\kappa$ is smaller than $\bar{\kappa}$ for $\theta<50''$; since $\kappa$ is proportional to the projected surface density, this reflects the fact that the substructure fluctations are fairly small in amplitude in relation to the mean density. Hence the difficulty in using convergence or shear to measure substructures; the signal-to-noise on $\sigma_\kappa$ will be smaller than that on $\bar{\kappa}$.

The upper and lower lines for $\bar{\kappa}$ and $\sigma_\kappa$ can be compared to see the effect of including or omitting substructure; the mean of the convergence is a little higher on scales $\ga 30''$ if we include substructure, and the standard deviation of the convergence is several times larger. This is to be expected; the presence of substructure changes the profile a little, and significantly alters the spatial variation of the matter distribution.

We now consider the 1-flexion galaxy-galaxy signals, again calculated in annuli with width $0.8''$, centred on the mode of the galaxy's $\kappa$ distribution, and smoothed with a $0.6''$ diameter top hat. Figure \ref{fig:fstats} shows the mean $\bar{F}$ and standard deviation $\sigma_F$ for the radial 1-flexion in annuli, for an ensemble of background galaxies behind the halos. We show results with either MEF filtering or no filtering, and with or without substructure, as it is important to understand the impact of our filter on the results. We should emphasize that this filter is applied to the N-body simulations to increase their realism; it is not a filter applied at the measurement stage.

We immediately see that the filtering of the N-body simulation is valuable; in the cases where no filtering is applied, the flexion standard deviation is substantial even in the absence of substructure. This is due to the particle noise in the simulations that leads to spurious high frequency convergence gradients. However, with the MEF filter in the absence of substructure, the flexion standard deviation is at least an order of magnitude smaller  than the unfiltered standard deviation.

Importantly, we see that when substructure is included (with MEF filtering of the simulations), $\sigma_F$ dominates over $\bar{F}$ on scales greater than about $10''$, and is two orders of magnitude larger than the no-substructure signal. The former observation confirms our claim that flexion variance is of interest, as it has much larger signal-to-noise than standard galaxy-galaxy flexion. The latter observation shows that it is substructure which is activating this signal; this is important, as the overall ellipticity of the central halo might have given a flexion variance \citep[c.f.][]{2009arXiv0903.3938H} but our result demonstrates that this is negligible for our statistic. To measure halo ellipticity with flexion, one should instead follow the methods of \citet{2009arXiv0903.3938H}.

The flexion variance also dies off more slowly with $\theta$ than $\bar{F}$ does. This is presumably due to the different phenomena being probed: $\bar{F}$ is probing the mean gradient of the overall galaxy halo density; while $\sigma_F$ is dominated by the gradients of substructure halos within the annuli, which remain similar regardless of which annulus they are found in. 

\subsection{Insensitivity to Varying Dominant Halo Mass}

\begin{figure}
\psfig{file=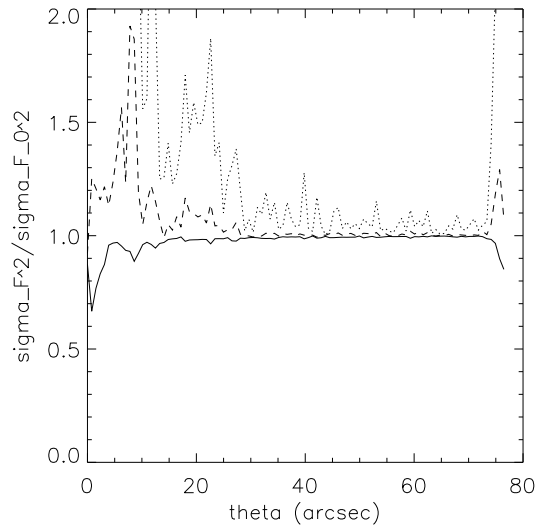,width=8cm}
\caption{Ratio of $\sigma_F^2$ with modified central halo amplitude to the usual $\sigma_F^2$. Solid: no central halo; dashed: total mass standard deviation of $4\times10^{12}M_\odot$; dotted: total mass standard deviation of $10^{13}M_\odot$.}
\label{fig:varym}
\end{figure}

An issue that we have not yet dealt with is the question of whether combining galaxies with various masses contributes a dominant unwanted signal to our galaxy-galaxy flexion signal. It is common in galaxy-galaxy lensing to divide up or scale one's foreground lens set according to a mass proxy such as luminosity \citep[e.g.][]{2006A&A...455..441K} or stellar mass \citep[e.g.][]{2006MNRAS.368..715M}. Nevertheless, within such a bin in mass, will the flexion variance signal be dominated by the overall mass variance rather than the substructure signal?

In order to test this, we measured the mean flexion signal for each of the main halos. We then subtracted this signal from each halo's flexion map, leaving the flexion due to substructure alone. We then optionally re-added a random proportion of the mean flexion signal, leading to a total mass standard deviation of $4\times10^{12}M_\odot$ or $10^{13}M_\odot$, and remeasured the ensemble flexion variance signal for our 12 realizations. The effects on our signal are shown in figure \ref{fig:varym}.

Firstly we notice that the difference between the shear variance with and without central halos is very small; most of the signal in our original ensemble is therefore coming from substructure. This is borne out by the case where we have a halo mass variance of $4\times10^{12} M_\odot$. The ratio of this signal to our original is within a few per cent of unity beyond $\theta\ga 20''$. In the extreme case where the mass variance is $10^{13}M_\odot$, we obtain about 10\% of the flexion variance from the main halos for $\theta>30''$, but this case corresponds to hardly having any mass binning at all.

We conclude, then, that provided reasonable steps are taken to deal with the galaxy-galaxy lensing in mass or luminosity bins, flexion variance is totally dominated by the substructure signal.

\subsection{Signal-to-Noise for Galaxy-Galaxy Flexion}

\label{sn}

The question now arises whether flexion variance is measurable with sufficient accuracy on the relevant scales to constrain substructure. The difficulty is that source galaxies are measured with a
substantial intrinsic flexion \citep[c.f.][]{2005ApJ...619..741G}; the distribution has a strong central peak and wide wings, such that the measured distribution has a 68\% range $\sigma_F\simeq0.1$ arcsec$^{-1}$. Clearly this intrinsic variance dominates over the substructure variance at all scales. However, following a statistical approach from \citet{2000MNRAS.318..625B}, one can estimate the flexion variance due to substructure, $\sigma^2_{sub}$, as an excess variance within an annulus, over and above that due to shape noise $\sigma^2_{intrinsic}$:

\begin{equation}
\sigma_{sub}^2 \simeq \sigma_{annulus}^2 - \sigma_{intrinsic}^2
\end{equation}
where $\sigma^2_{annulus}$ is the measured total variance within an annulus, while $\sigma^2_{intrinsic}$ is the variance measured for the whole ensemble of source galaxies. Since this equation is only exact for Gaussian distributions, we have checked that it is approximately true for our simulation flexion distribution and realistic shape noise distribution; we find that $\sigma^2_{sub}$ is correctly estimated to within $5\%$.  As described by \citet{2000MNRAS.318..625B}, the error on this estimator is approximately

\begin{equation}
\sigma[\sigma_{sub}^2] \simeq \sigma_{intrinsic}^2/\sqrt{N}
\label{eqn:sigonrootn}
\end{equation}
where $N$ is the number of objects in the annulus. 

\begin{figure}
\psfig{file=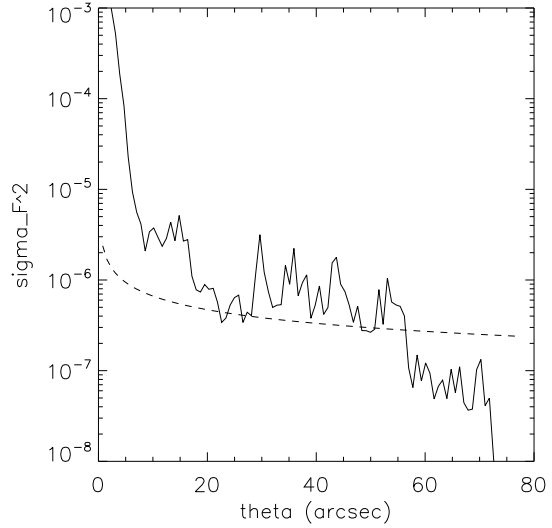,width=8cm}
\caption{Flexion variance estimator (solid line) in units of (arcsec)$^{-2}$, and expected noise level in each annulus (dashed line) for the Euclid survey described in the text.}
\label{fig:estnoise}
\end{figure}

We can now use this equation to estimate how strong a signal we expect for a Euclid-like survey. We use $\sigma_{F int}=0.1$, and calculate $N$ using the standard survey parameters of \citet{2007MNRAS.381.1018A}. We examine the case of choosing 9 foreground galaxies per sq arcmin with $z_{med}=0.46$, 26 background galaxies per sq arcmin with $z_{med}=1.1$, and a survey area of 20000 sq deg. Figure \ref{fig:estnoise} shows the expected value of our estimator $\sigma_{F sub}^2$ (solid line) and the noise level on this estimator in each annulus (dashed line). Note that for a large range in $\theta$, the S/N is considerable: $S/N\simeq1$ to 5 for each of 50 annuli between $20''$ and $60''$. This suggests that substructure can be studied in great detail using this method.

\subsection{Sensitivity to Substructure Content}

\begin{figure}
\psfig{file=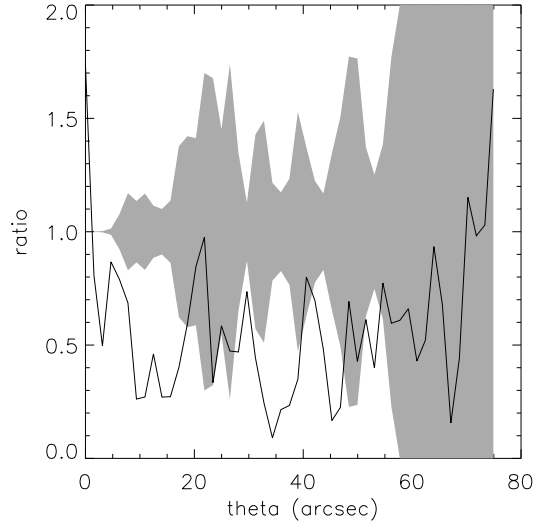,width=8cm}
\caption{Ratio of flexion variances with substructure and without 50\% of the subhalos. Grey shading shows region within $1\sigma$ uncertainty of flexion variance with substructure.}
\label{fig:half}
\end{figure}

\begin{figure}
\psfig{file=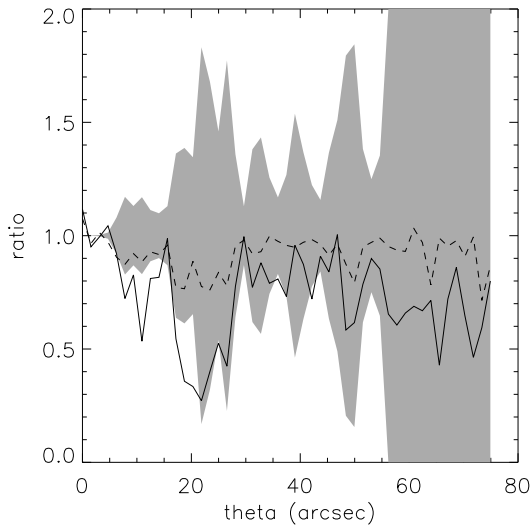,width=8cm}
\caption{Ratio between (a) flexion variances with all subhalos, and with all subhalos $>5\times10^8 M_\odot$ (dashed line); (b) flexion variances with all subhalos, and with all subhalos $>10^9 M_\odot$ (solid line). Grey shading shows region within $1\sigma$ uncertainty of flexion variance with all subhalos.}
\label{fig:lt1e9}
\end{figure}

To pursue this point, we provide some examples of the degree to which we can distinguish between different substructure scenarios.

First, we examine the impact of removing half of the subhalos at random in each of our simulations. We carry out the analysis of sections \ref{signal} and \ref{sn} for these modified simulations, and show how their flexion variance signal differs from the usual case on Figure \ref{fig:half}. Here we have used 50 bins between 0 and 80$''$ for clarity. 

We note that the flexion variance is itself approximately halved in this case. A $\chi^2$ measure for the significance of this difference is the sum of $(\sigma_F^2 - \sigma_{F half}^2)^2/\sigma[\sigma^2_{sub}]^2$ for all annuli, and is in this case $\simeq250$, confirming that such a substructure configuration would be strongly distinguished.

However, that scenario is not expected physically. More plausibly, if dark matter has a non-negligible temperature, this will preferentially remove substructures up to a certain mass threshold. To examine whether this type of phenomenon would be detectable by flexion variance, we remove all halos with mass $<5\times10^8M_\odot$ or $<10^9M_\odot$ from our simulations. We again carry out the analysis of sections \ref{signal} and \ref{sn} for the modified simulations. The results of this process are shown on Figure \ref{fig:lt1e9}.

We see that there is only a small difference between the signals for all subhalos and for all subhalos $>5\times10^8M_\odot$; they are only slightly distinguishable within the error expected as shown by grey shading, with $\chi^2=4.3$. Care should be taken in drawing conclusions from this, as halos with mass $5\times10^8M_\odot$ are still near our resolution limit; we will explore lower mass substructure in future work. On the other hand, the difference between the $>10^9M_\odot$ case and the full substructure case is easily detected, with $\chi^2=37$. According to the thresholds given by \citet{2001ApJ...558..482B}, this would approximate detection of a 2keV mass scale for warm dark matter.

\section{Conclusions}
\label{Conclusions}

In this paper we have explored the utility of galaxy-galaxy flexion variance for the purpose of measuring the degree of substructure on galactic scales.

We have described the relevant statistics, showing how flexion radial and systematic modes are constructed in a galaxy-galaxy lensing context, and introducing the concept of flexion variance in annuli. We have explained how this is a more suitable probe than galaxy-galaxy-galaxy flexion (or flexion correlation functions in annuli), as the latter has a very rapid drop in signal as a function of angle.

We have gone on to test the use of these statistics by adopting a set of N-body simulations of galaxy halos, including substructure, or removing this substructure by means of halo-finding algorithms. We have then calculated the weak lensing convergence associated with these halos, applying the suitable MEF filter to reduce the effect of particle shot noise.

We have shown that the flexion variance is able to detect substructures on a halo by halo basis. This carries through to the full galaxy-galaxy flexion variance expected with an ensemble of halos; the signal found is substantially larger than that for the flexion mean, or indeed the flexion variance in the absence of substructures. The underlying central halo mass variance is not found to be a dominant source of noise for this signal.

We have made predictions for the level of the flexion variance signal-to-noise for a space-based survey such as Euclid, finding that substructure amplitudes will be measured with significant precision. This allows us to discriminate between substructure scenarios with different numbers and masses of halos, and will enable a constraint on dark matter particle mass.

\section{Acknowledgements}

We would like to thank J\"{u}rg Diemand and Joachim Stadel for making their simulation output available to us. We would also like to thank Rob Crittenden, Tom Kitching and Barnaby Rowe for very useful discussions. DB is supported by an STFC Advanced Fellowship and RCUK Research Fellowship. AA is supported by the Zwicky Fellowship at ETH Zurich.

\bibliographystyle{mn2e}
\bibliography{flex}

\end{document}